# Metal oxide decoration on Si-FETs for selective gas sensing at room temperature

Areej Shahid
*Department of Electrical and Electronic Engineering, The University of Melbourne, Parkville, Victoria, Australia*
Email: areejs@student.unimelb.edu.au

Ranjith R Unnithan
*Department of Electrical and Electronic Engineering, The University of Melbourne, Parkville, Victoria, Australia*
Email: r.ranjith@unimelb.edu.au

Jackson Gum
*Department of Materials Science and Engineering, Monash University*
Clayton, Victoria 3800, Australia;
Email: Jackson.Gum@monash.edu

Sudha Mokkapati
*Department of Materials Science and Engineering, Monash University*
Clayton, Victoria 3800, Australia;
Email: sudha.mokkapati@monash.edu

*Abstract–Metal oxide semiconductors have been thoroughly studied for gas sensing applications due to the electrical transduction phenomenon in the presence of gaseous analytes. The chemiresistive sensors prevalent in the applications have several challenges associated with them inclusive of instability, longevity, temperature/humidity sensitivity, and power consumption due to the need of heaters. Herein, we present a silicon field effect transistor-based gas sensor functionalized with CuO. The oxidized Cu thin film acts as a selective room temperature $H_2S$ sensor with impressive response and recovery. Using this methodology, we propose a standalone compact enose based on our results for a wide spectrum of gas detection.*

*Keywords— FET gas sensor, Si, metal oxide semiconductor, copper oxide*

## 1. Introduction

Due to semiconductor properties and transducer characteristics, several metal oxides have been investigated as gas sensors such as $ZnO$, $TiO_2$, $SnO/SnO_2$, $WO_3$, $CuO/Cu_2O$, and $V_2O_5$ [1]. Based on the type of gases (reducing or oxidizing), there might be an increase or decrease in the resistance of the sensing material. Copper oxide (CuO), specifically, is a p-type semiconductor and detects the decrease in conductivity upon exposure to reducing gases such as $H_2S$ or $NH_3$, or an increase in conductance upon interacting with oxidizing gases such as $CO_2$, $NO_2$, etc. [2]. However, most gas sensors work on chemiresistive techniques and require electric heaters to operate in ambient environments, consuming a lot of power. Also, the resistive structure is very bulky and comparatively less sensitive to detect minimal changes in analyte concentrations [3]. Other types include optical and electrochemical gas sensors which have portability and short lifetime issues.

Therefore, field effect transistor (FET) type gas sensors are emerging as a solution to the aforementioned challenges associated with contemporary gas detection techniques. FETs can be miniaturized and calibrated to detect temperature, humidity, or chemical analytes with high precision and signal gain [4]. Herein, we present a chemical detector based on silicon FET, functionalized with CuO to study the sensing of three toxic gases $H_2S$, $NH_3$, and $NO_2$. The fabricated device is electrically characterized and exposed to target gases using microfluidic channels integrated on the sensor surface to measure changes in electrical resistance. The results confirm the redox reactions between the CuO sensing layer and the target analytes at room temperature. The study opens avenues to explore more sensing layers and channel transducers to selectively detect (identify) amidst co-existing gases.

## 2. Materials and Methods

The FETs were fabricated using standard micro and nanofabrication techniques including evaporation, lithography, and lift-off processes [5-7]. A 0.7mm thick Si substrate coated with 200nm $SiO_2$ is diced into 15mm*20mm substrates precleaned using acetone, IPA, and DI water. A 250 μm x 200 μm channel lithography was done for 5 tandem FETs on the substrate using negative resist AZ2035. This was followed by 30nm Si deposition using the E-beam evaporation. Lithography and deposition were done for a 70μm *100μm area of the sensing layer. A thin film of $Cu_{6nm}$ was deposited using e-beam evaporation followed by oxidation in a furnace to make CuO nanoparticles. Finally, 100nm Au electrodes, adhering via 7nm Ti were deposited using similar lithography and evaporation parameters. The fabrication process flow, top view, and side view of the CuO functionalized Si-FETs are shown in Fig 1. (a-c). Finally, to test the fabricated sensors with target gas, a microfluidic channel was fabricated and placed on the sensing layer of the tandem FETs. The devices were electrically tested for IV characteristics and then exposed to the target gases for response analysis.

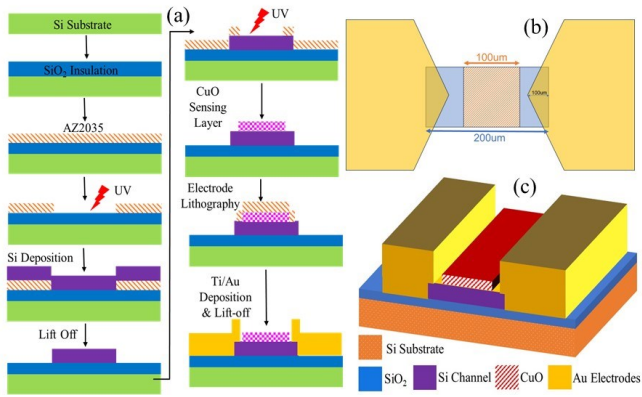

Fig 1. (a) Fabrication process flow of the CuO decorated Si-FET, (b) top view and dimension of FET, (c) Side view of FET

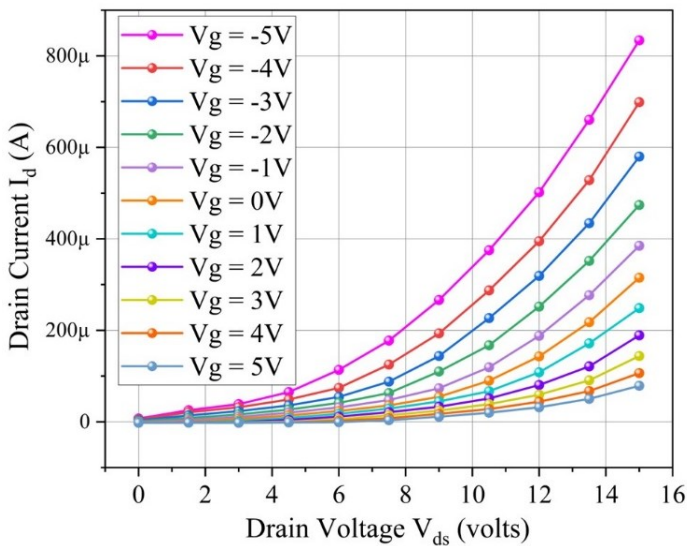

Fig 2. Forward characteristic drain current ($I_{ds}$) and drain-source voltage ($V_{ds}$) curves for Si FETs functionalized with CuO

## 3. RESULTS AND DISCUSSION

### A. Electrical Characteristics

The devices were electrically tested for IV characteristics using Agilent technologies sourcemeter and QuickIV software. The connections to source, drain, and back gate were established and a sweeping voltage of -15 to +15 volts was provided across source-drain ($V_{ds}$). The gate voltage ($V_g$) varied from -5 to +5V and was used as a stepper. The forward characteristics of CuO-decorated Si-FET are shown in Fig 2.

### B. $H_2S$, $NH_3$ and $NO_2$ Detection

The PDMS channel adhered on top of the sensing material forming a microfluidic channel to pass the target gases. The CuO sensing layer works on the oxide ions adsorption principle. In pure air, the environmental oxygen molecules adsorb onto the CuO surface forming the oxide ions $O_2^-$, which either oxidize or reduce the target analyte based on the reducibility capacity in comparison to CuO. Equal concentrations (50ppm) of $H_2S$, $NH_3$, and $NO_2$ were passed through the sensor track and allowed to stay on top of the sensing layer. The sensor was tested for the whole spectrum of bias conditions. The response to the gases is shown in Fig 3 where the left y-axis shows drain current $I_d$ and the right y-axis shows the resistance R. Fig 3(a) and (b) show the response to two reducing gases, $H_2S$ and $NH_3$, respectively.

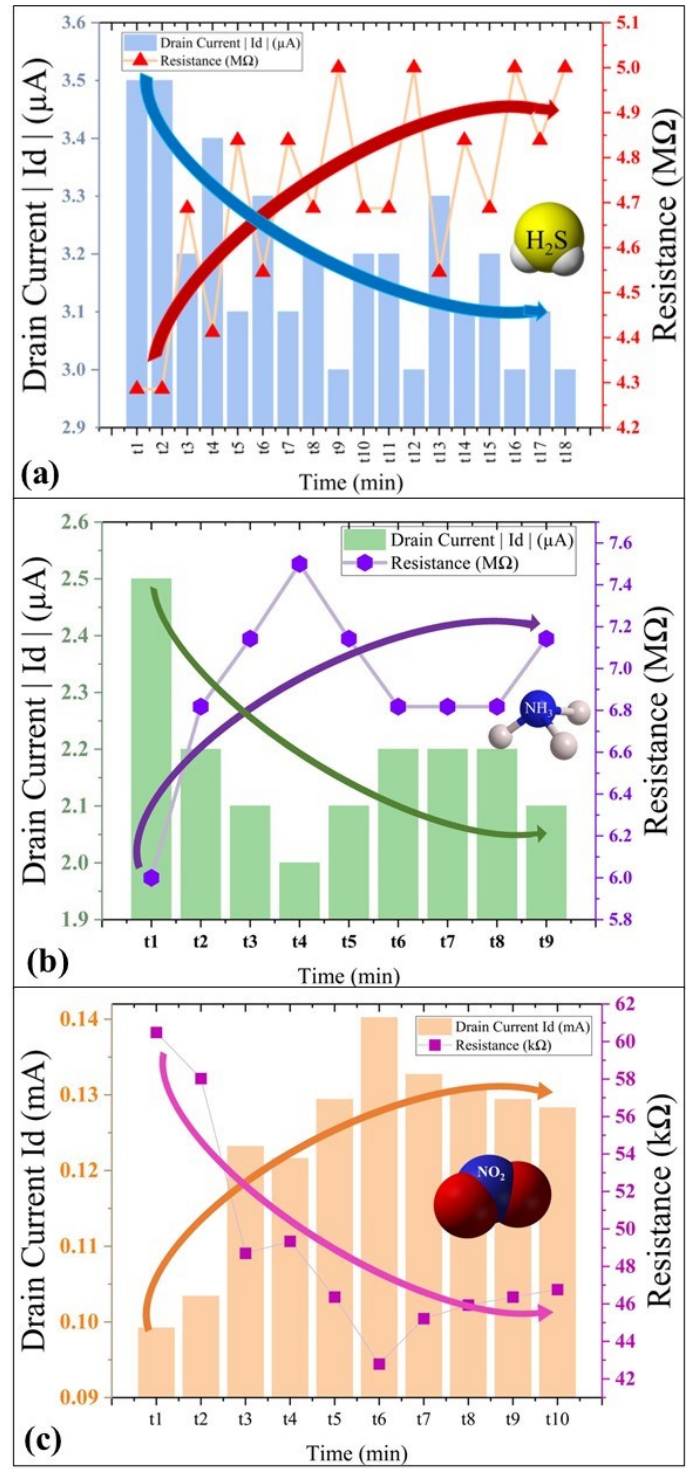

Fig 3. Detection results of CuO functionalized Si-FETs in terms of drain current $I_d$ and resistance R upon exposure to (a) $H_2S$, (b) $NH_3$, and (c) $NO_2$

As CuO is p-type in nature, therefore, its interaction with the reducing gases results in an increase in the sensing film resistance (with a corresponding decrease in the drain current) as shown in Fig 3(a and b). However, it is noticed that the proportional response to $H_2S$ is higher than that of $NH_3$, which makes it more selective to this gas. This is in line with the higher reducibility rates of $H_2S$ gas [8,9].

In contrast, the $NO_2$ response is reciprocal to that of the above-mentioned reducing gases. Upon exposure to this gas, the oxide ions interact with the gas molecules and free the electrons into the CuO sensing layer, which narrows down the depletion region. Resultingly, we can observe an increase in the current and a decrease in the resistance (Fig 3c). It is to be noted here that for $NO_2$ sensing, higher and inverse bias conditions were used, which are reflected in higher currents. However, since $NO_2$ detection is in opposite polarity, the sensor can be investigated further as a multiplexing switch to detect multiple gases [10].

4. CONCLUSION

In this study, we present the FET-type gas sensor using silicon as a transducer and lab-synthesized CuO as a selective gas sensing material. High detection levels were observed for $H_2S$, $NH_3$, and $NO_2$ at room temperature for 50 ppm concentrations. We noticed an increase in resistance for the oxidizing gases $H_2S/NH_3$ and a decrease in resistance for the oxidizing gas $NO_2$. The higher sensitivity to $H_2S$ as compared to $NH_3$ and also an inverse polarity to that of $NO_2$ (detection zone) is an attribute of great interest for selective gas sensing and identification purposes. This study forms the basis for our future investigation of metal oxides in FET architectures as ultrasensitive and selective room temperature gas sensors. The multi-gas sensing and selectivity using particular voltage bias conditions can be utilized to implement a standalone enose on a single chip.

ACKNOWLEDGMENT

This work was performed in part at the Melbourne Centre for Nanofabrication (MCN) in the Victorian Node of the Australian National Fabrication Facility (ANFF).